\documentstyle[prl,aps,epsfig]{revtex} 

\begin{document}
\def\lambdabar{{\mathchar'26\mkern-9mu\lambda}}

\draft
 

 \twocolumn[\hsize\textwidth\columnwidth\hsize  
 \csname @twocolumnfalse\endcsname              

\title{Descritpion of Exotic Nuclei  Using Continuum Shell Model} 

\author{N. Michel\dag,  J. Oko{\l}owicz\dag\ddag ~and M. P{\l}oszajczak\dag}
\address{\dag\ Grand Acc\'{e}l\'{e}rateur National d'Ions Lourds, 
CEA/DSM -- CNRS/IN2P3, BP 55027, F-14076 Caen Cedex 05, France}
\address{\ddag\ Institute of Nuclear Physics, Radzikowskiego 152, 
PL - 31342 Krakow, Poland}

\date{today}

\maketitle

\begin{abstract}
In weakly bound exotic nuclei,
number of excited bound states or narrow resonances
is small and, moreover, they couple strongly to the particle continuum.
Hence, these systems should be described in
the quantum open system formalism which does not artificially separate
the subspaces of (quasi-) bound and scattering states. The 
Shell Model Embedded in the Continuum provides a novel approach which 
solves this problem. Examples of application in $sd$ shell nuclei will be
presented. 
\end{abstract}

\bigskip
 
 ]  

\narrowtext
\section{Introduction}
 A realistic account of the low-lying states properties in exotic nuclei
requires taking into account the coupling between discrete and continuum states. 
This aspect is particularly important in studies near the drip line where 
one has to use both
structure and reaction data to understand basic properties 
of these nuclei. Within the Shell Model Embedded in the 
Continuum (SMEC) \cite{bnop1}, one may obtain a unified description of the 
divergent characteristics, such as the spectra (energies of states, transition
probabilities, proton/neutron emission widths, $\beta -$decays, etc.) and
the reactions involving one-nucleon in the continuum 
(proton/neutron capture processes,
Coulomb dissociation reactions, elastic/inelastic proton/neutron reactions,
etc.). The accumulation of divergent observables analyzed in the same
theoretical framework provides a stringent
test of the effective interactions and
permits to asses the mutual complementarity of reaction and structure data
for the understanding of exotic configurations and decays in those weakly bound
nuclei.
 
In the SMEC formalism, the subspaces of (quasi-) bound (the $Q$ subspace) and scattering (the
$P$ subspace) states are separated using the projection operator technique
\cite{bartz3}. $P$ subspace contains asymptotic channels, made of
$(N-1)$-particle localized states 
and one nucleon in the scattering state. $Q$ subspace contains 
many-body localized states which are
build up by both the bound state single-particle (s.p.) wave functions
and the s.p.\ resonance wave functions. 
The wave functions in $Q$ and $P$ are properly renormalized
in order to ensure the orthogonality of wave functions in both subspaces. The
details of the approach can be found in \cite{bnop1,ostatnia}. 

The salient feature of SMEC is that
the (quasi-) bound many-body states in $Q$ are given by 
the multiconfigurational Shell Model (SM) with the realistic
effective interaction, providing the internal mixing of configurations. The coupling between bound and scattering
states is described by 
the density dependent interaction (DDSM1) \cite{ostatnia,plb}.
This interaction provides an external mixing of configurations via the virtual excitations of
particles to the continuum states. A subtle balance of external and internal
configuration mixing explains energies and widths of levels, $(p,p^{'}$)
excitation functions, radiative capture processes, {\it etc.}.

To generate radial s.p.\ wave functions in $Q$ 
and the scattering wave functions in $P$, as a first guess,
we use the potential of Woods-Saxon (WS) type
with the spin-orbit : $ V_{SO} {\lambdabar}_{\pi}^2 (2{\bf l}\cdot{\bf s})
r^{-1}d{f}(r)/dr$, and Coulomb parts included.  
${\lambdabar}_{\pi}^2 = 2\,$fm$^2$ is the pion Compton wavelength and
$f(r)$ is the spherically symmetrical WS formfactor.
The Coulomb potential $V_C$ is calculated for a uniformly charged sphere.
This 'first guess' potential $U(r)$, is then self-consistently modified by the residual interaction. 
The iterative procedure yields the self-consistent
potential $U^{(sc)}(r)$, which depends on channel angular momentum and differs significantly 
from the initial potential in the interior of the potential well. For weakly bound configurations, 
also the surface properties of initial potential are changed by the residual
coupling. As a result of this self-consistent
coupling, the initial SM Hamiltonian in $Q$ becomes energy dependent and its eigenvalues become complex for
decaying many-body states.

\section{The first-forbidden mirror $\beta$ decays in A=17 nuclei}
 First-forbidden $\beta^{+}$ decay rate ($f^+$) from the 
ground state (g.s.) $J^{\pi}=1/2_1^-$ of $^{17}$Ne to the weakly bound
state $J^{\pi}=1/2_1^+$ in $^{17}$F exhibits an abnormal asymmetry  
with respect to its mirror $\beta^{-}$ decay rate ($f^-$) 
of $^{17}$N into a well bound excited state of $^{17}$O \cite{borge}.
This asymmetry has been explained either by large
asymmetry of radial sizes of $1s_{1/2}$ s.p. orbits involved in bound states of
$^{17}$F/$^{17}$O and $^{17}$Ne/$^{17}$N \cite{borge} or by the charge-dependent
effects leading to different amplitudes of 
$\pi(0p_{1/2}^21s_{1/2}^2)\nu(0p_{1/2}^{1})$
and $\nu(0p_{1/2}^21s_{1/2}^2)\pi(0p_{1/2}^{1})$ components in
the g.s. wave functions of $^{17}$Ne and $^{17}$N, 
respectively \cite{millener}. These two qualitative analysis of the mirror
asymmetry lack a  consistent treatment of radial properties of
the wave functions which are involved in these transitions. 

\begin{figure}
\epsfig{figure=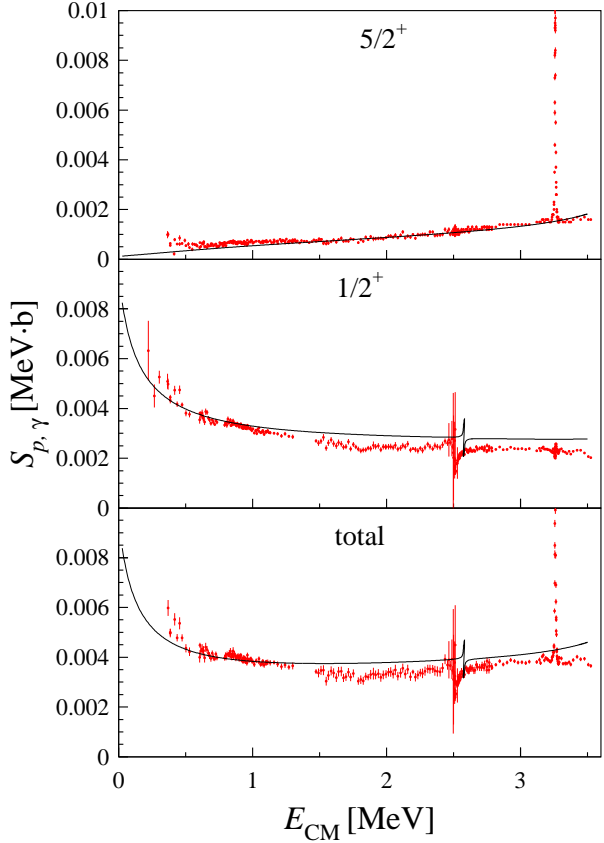,height=12cm}
\caption{The astrophysical $S$-factor for the reaction
$^{16}\mbox{O}(p,\gamma)^{17}\mbox{F}$ leading to the states 
$5/2_{1}^{+}$ and $1/2_{1}^{+}$ is plotted
as a function of the center of mass energy $E_{CM}$ \protect\cite{plb}. 
The experimental data are from \protect\cite{morlock}. 
The diffusenes parameter of the initial WS potential 
$U(r)$ is $a=0.55$ fm. The depth of  $U(r)$ is chosen in such a way that the
self-consistent potential yields the binding energies of proton
s.p. orbits $0d_{5/2}$ and $1s_{1/2}$ at the 
experimental binding energies of $5/2_{1}^{+}$ and $1/2_{1}^{+}$ many-body 
states  in $^{17}$F (from \protect\cite{plb}). 
}
\end{figure}
The stringent constraint on the diffuseness of the mean-field generating radial s.p. wave functions is provided by the 
proton radiative capture cross section (the astrophysical $S-$factor) 
$^{16}$O$(p,\gamma )^{17}$F to the 'proton halo state'
$1/2_1^+$, which in the formalism of SMEC can be calculated using not only the same wave
functions in $Q$ and $P$, but also the same SM interaction, residual coupling 
or the initial average potential, as those used  in the calculation of the
decay rate $^{17}$Ne($\beta^+$)$^{17}$F and the spectrum of $^{17}$F. Fig. 1
compares measured astrophysical $S-$factor for the reaction 
$^{16}$O$(p,\gamma )^{17}$F with the results of the SMEC calculations using ZBM
interaction in $(p_{1/2}d_{5/2}s_{1/2})$ shells. The capture cross-section data
can be well reproduced using $U(r)$ with the surface diffuseness in the range : $a=0.55\pm 0.05$ fm. 

Having determined the surface features of the mean-field,
we can estimate the charge-dependent effects in $1s_{1/2} \to 0p_{1/2}$
dominant contribution to the first-forbidden mirror $\beta$ transitions 
systems. The calculation of the nuclear matrix elements and, hence, 
the first-forbidden $\beta -$decay rates are done using
the formalism of Towner and Hardy \cite{towner}. Following the analysis of
Warburton et al. \cite{war94} for $A\sim 16$, the matrix element 
$\xi^{'}v$ of the time-like piece of
the axial current is multiplied by a constant factor 1.61
to account for an enhancement due to meson-exchange currents. 

The experimental rate  $f^+$ for $^{17}$Ne(${\beta^+}$)$^{17}$F 
is reproduced by the SMEC calculation
with ZBM-F interaction. This hybrid interaction 
reproduces exactly the experimental 
energy splitting between the g.s. $5/2_1^{+}$ and the first excited state 
$1/2_1^{+}$  in $^{17}$F. However, for the same interaction the measured rate $f^{-}$
for $^{17}$N($\beta^-$)$^{17}$O is overpredicted by a factor $\sim 3$.
Since the radial dependences which
are consistent with the proton capture data give an excellent fit of
both the $\beta^+$ decay rate and the spectrum of $^{17}$F, 
the discrepancy for $f^-$ and $f^+/f^-$ is uniquely due to the deficiency of ZBM-F
interaction to reproduce the configuration mixing in $^{17}$O and $^{17}$N, {\it
i.e.} by the charge-dependent effects in the $Q$ space SM Hamiltonian. 

How large are these effects?  An essential parameter here is the
amplitude of a component $(1s_{1/2}^20p_{1/2}^{-1})$ in the g.s.
wave functions of $^{17}$Ne and $^{17}$N. Experimental values for $f^-$ and
$f^+/f^-$ are reproduced if the amplitude of
$(1s_{1/2}^20p_{1/2}^{-1})$ in g.s. of $^{17}$N is reduced
by $\sim 30 \%$. In a simplest way, this can be achieved
by reducing the separation of $0d_{5/2}$
and $1s_{1/2}$ shells in ZBM-F interaction. This new interaction, called
ZBM-O*, reproduces also well the experimental spacing of $5/2_1^{+}$ and
$1/2_1^{+}$ levels in $^{17}$O. One should notice that this important
reduction of the $(1s_{1/2}^20p_{1/2}^{-1})$ amplitude concerns a very small
component of the g.s. wave function of $^{17}$Ne and $^{17}$N. The dominant
component, which is $(0d_{5/2}^20p_{1/2}^{-1})$, changes by less than $5 \%$
going from ZBM-F interaction to ZBM-O* interaction. Also the dominant 1p-0h configuration 
in g.s. $5/2_1^+$ and first excited $1/2_1^+$ states of $^{17}$F (ZBM-F) 
and $^{17}$O (ZBM-O*) is only slightly modified ($\sim 12 \%$). 
Of course, one should be aware that
these estimates of charge-dependent effects in SM Hamiltonian
can be somewhat affected by the chosen ZBM valence space. In particular, the absence of $0p_{3/2}$ and
$0d_{3/2}$ subshells leads to an amplification of
the sensitivity in the $1s_{1/2} \rightarrow 0p_{1/2}$ contribution to the
charge-dependent effects \cite{millener}. Nevertheless, for the first time the
quantitative extraction of charge-dependent effects, separately 
on the radial properties of
wave functions and on the configuration mixing in mirror systems, became 
possible.

\section{Binding energies in neutron-rich oxygen isotopes}
 As a second example of application of the SMEC formalism, we shall consider the
correction to the SM g.s. energy due to the continuum coupling.
In this section, 
\begin{figure}
\epsfig{figure=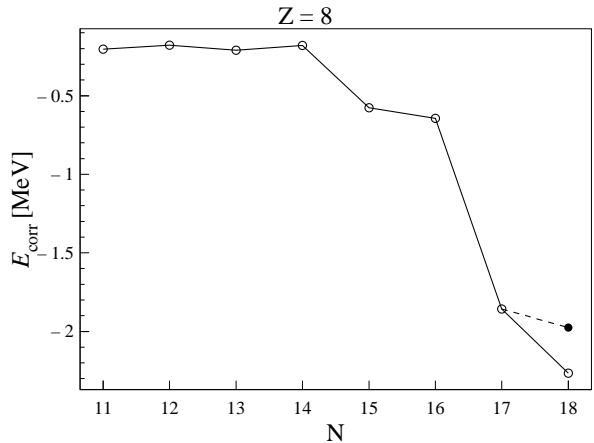,height=6cm}
\caption{The SMEC energy correction to the g.s. masses of oxygen isotopes
is calculated in the full $sd$ shell using a Widenthal effective interaction
\protect\cite{wildenthal}.
}
\end{figure}
we use the valence
space of full $sd$ shell and the effective interaction of Wildenthal
\cite{wildenthal}. One- and two-neutron separation energies ($S_n$ and
$S_{2n}$, respectively)
have been calculated for this isotope chain using SM in the same effective space
and for the same effective interaction
\cite{caurier}. Fig. 2 shows the above mentioned energy correction 
$E_{corr}$ to the SM masses in the
chain of neutron-rich isotopes of oxygen. $E_{corr}$ 
strongly increases while approaching the neutron drip line. The
'odd-even staggering' reflects a sensitive dependence of this
correction on the one-neutron separation energy $S_n$, which exhibits a similar
dependence. This alone is however not sufficient. The second important effect 
is related to odd-even variation in the density of many-body 
states in the $N-1$ system determining the overall number of channels by which
system $N$ couples to the scattering continuum.

Particularly large 
coupling matrix elements correspond to the isoscalar couplings 
between proton and neutron fluids. In the example shown in Fig. 2, these 
couplings are absent (number of protons in $sd$ shell equals zero), 
but one expects that the SMEC energy correction to SM masses 
will strongly increase when both fluids are present.
 This may indicate a large shift of the
position of the neutron drip line between oxygen and fluor. 

SMEC correction to SM masses depends strongly on the radial wave functions
in $Q$. An example of this kind can be seen in Fig. 2. For oxygen isotopes, 
the standard isospin dependence of the depth of the central potential
yields $0d_{3/2}$ s.p. state for neutrons unbound for $N<18$. With
increasing $N$, the energy of $0d_{3/2}$ resonance is going down 
and for $N=18$ it becomes
bound by $\sim 200$ keV. In view of the uncertainty concerning the isospin
dependence of the central potential, we have calculated
$E_{corr}$ for $^{26}$O ($N=18$) by keeping 
$0d_{3/2}$ in the continuum at $\sim 50$ keV. The full point for 
$N=18$ gives an idea of the sensitivity of
$E_{corr}$ to the asymptotic properties of $0d_{3/2}$ s.p. state.

\begin{figure}[t]
\epsfig{figure=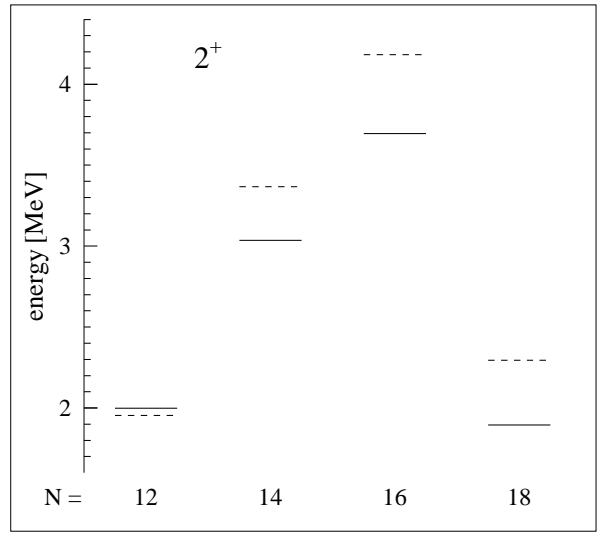,height=7cm}
\vskip 1truecm
\caption{Excitation energy of $2_1^+$ excited state in even-$N$ isotopes of O
calculated in SMEC (solid line) and in SM (the dashed line).
}
\end{figure}
\begin{figure}[t]
\epsfig{figure=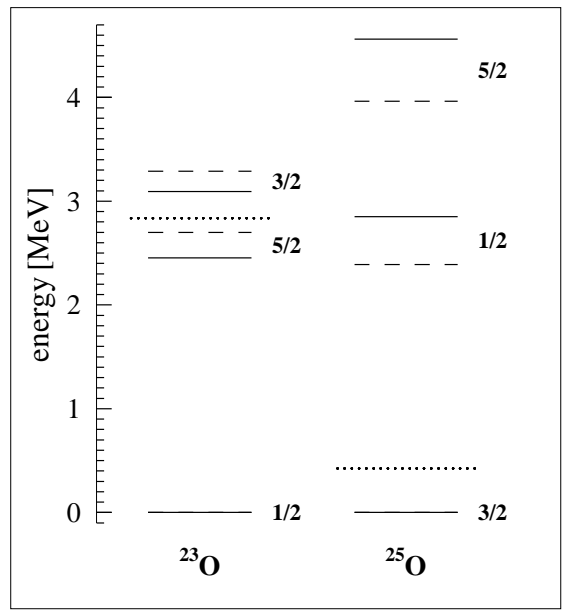,height=8cm}
\vskip 1truecm
\caption{Lowest energy states in odd-$N$ oxygen isotopes around $^{24}$O. 
The dotted line shows the position of one-neutron emission threshold. SMEC and
SM results are shown with the solid and dashed lines, respectively.
}
\end{figure}
When approaching neutron drip line, the two neutron separation energy
$S_{2n}$ decreases fast and may become smaller than $S_n$. 
In this case, the dominant continuum coupling corresponds to a virtual
excitation of nucleon pairs, which are not yet included in SMEC. For that reason,
the SMEC correction close to neutron drip line has to be considered as 
a lower limit of the continuum influence on SM masses.

Independent information about the configuration dependence of the 
continuum coupling is provided by the position
of the $2_1^+$ state in even-$N$ oxygen isotopes. 
In Fig. 3, we compare results of SM and SMEC, which are obtained using the same
model space and interaction in $Q$. One can see that the relative
shift of $2_1^+$ with respect to the g.s. $0_1^+$ depends on $N$, and in some
cases the continuum coupling shifts this state 
upwards with respect to the SM prediction.
However particularly interesting is $^{24}$O, for which the dominant
g.s. configuration does not contain the contribution from $0d_{3/2}$
s.p. resonance, whereas this resonance is strongly populated 
in $2_1^+$ state. We can see that the 
relative shift of $2_1^+$ state in SMEC calculations for $^{24}$O
is particularly strong, reflecting different radial asymptotic properties of
the dominant configuration. 

A spectacular example of the dependence of the SMEC energy correction
on the s.p. radial wave functions in $Q$ is shown in Fig. 4. Here the lowest energy
states of $^{23}$O and $^{25}$O are shown for SMEC (the solid lines) and SM
(the dashed lines). Zero of the energy scale corresponds to the g.s. in both 
cases. G.s. of $^{23}$O does not contain any significant
component involving the occupation of the s.p. resonans $0d_{3/2}$,
and the SMEC correction for excited states $5/2_1^+$
and $3/2_1^+$  close to the threshold (the dotted
line) are larger than in the g.s. In $^{25}$O, 
on the contrary, $0d_{3/2}$ s.p. resonans is strongly occupied in the g.s. and
, therefore, excited states $1/2_1^+$ and $5/2_1^+$ in SMEC calculations 
are shifted upwards with respect to their position in the SM. 

Not only the absolute energy
corrections are large for nuclei near the neutron drip-line, as shown in Fig.
2 for the g.s. of even-even nuclei, but also the relative shifts 
of excited states with respect to the g.s. can be large and of different signs
depending on the radial features of s.p. orbits involved, the internal
configuration mixing provided by the SM effective interaction, and the external
configuration mixing which depends on the size of the coupling matrix elements
for different channels in $[(N-1)\otimes n]^{J^{\pi}}$. Systematic experimental
investigation of these salient tendencies of the multiconfigurational continuum
shell model remains a challenge for future studies.

\section{Outlook}
With the SMEC, the new paradigm is born for the 
microscopic understanding of low-lying
properties of unstable nuclei. Fruitful exchanges with the
standard SM, which nowadays experience a revival in medium-heavy nuclei and
allows to perform  no-core calculations till $^{16}$O, allows to hope for a new
wave of exciting future developments in the field of exotic nuclei.

\vskip 0.5 truecm
We wish to thank G. Mart\'{\i}nez-Pinedo, T. Otsuka and F. Nowacki for useful discussions.

\end{document}